\documentclass[journal,9pt]{IEEEtran}
\usepackage{cite}
\usepackage{amsmath,amssymb,amsfonts,bm}
\usepackage{algorithmic}
\usepackage{graphicx}
\usepackage{textcomp}
\usepackage{graphics}
\usepackage{graphicx}
\usepackage{epsfig}
\usepackage{amssymb}
\usepackage{amsthm}
\usepackage{times} 
\usepackage{amsmath} 
\usepackage{amssymb}  
\usepackage{graphics}
\usepackage{graphicx}
\usepackage{epstopdf}
\usepackage{lineno}
\usepackage{array}
\usepackage[numbers]{natbib}
\usepackage{mathtools}
\newcommand{\R}{\mathbb{R}}
\newcommand{\norm}[1]{\left\lVert #1 \right\rVert}
\def\BibTeX{{\rm B\kern-.05em{\sc i\kern-.025em b}\kern-.08em
    T\kern-.1667em\lower.7ex\hbox{E}\kern-.125emX}}
\usepackage{url}
\begin{document}
\title{Power-efficient Sampling Time and Resource Allocation in Cyber Physical Systems with Industrial Application}
\author{Atefeh. Termehchi, and Mehdi. Rasti
\thanks{Atefeh. Termehchi and Mehdi. Rasti are with the Department of Computer Engineering and Information Technology,
Amirkabir University of Technology, Tehran, Iran (e-mail: {atefetermehchy, rasti}@aut.ac.ir). }}

\maketitle

\begin{abstract}
Cyber Physical Systems (CPSs) are the result of convergence of computation, networking, and control of physical process. In this paper, we consider an industrial CPS consisting of several control plants and Rate Constrained (RC) users that communicate via a single cell OFDMA network. The problem of jointly determining the sampling instant of each control plant and allocating power and sub-carrier, in the CPS, is formulated. The problem is a multi objective optimization with aims of determining the next maximum allowable sampling instant of each control plant and minimizing the power consumption in uplink and downlink, considering the dynamics and desired performance of control plants, the quality of service requirement of RC users, power and sub-carrier constraints. To solve the multi objective optimization problem, a novel approach is proposed which decomposes the optimization problem into two smaller loosely coupled problems. we show the effectiveness of the proposed approach through simulation results.
 \end{abstract}

\newtheorem {theorem}{Theorem}
\newtheorem {lemma}{Lemma}
\newtheorem {corollary}{Corollary}
\newtheorem {remark}{Remark}
\newtheorem {assumption}{Assumption}
\vspace{10pt}
\section{Introduction}
\vspace{10pt}
\label{sec:introduction}
Cyber Physical System is the result of integration of three subsystems including computation, networking and control of physical process. CPSs are an integral part of many applications such as healthcare industry, smart grid, industrial automation, and avionic. With the introduction of CPS and Internet of Things (IoT) concepts in the industrial automation, the fourth industrial revolution, namely, Industry 4.0, is emerging \cite{Wollschlaeger2017TheFO}. The new technologies are expected to make a tremendous change in future industrial systems \cite{meng, emerging}. In the direction of designing Industry 4.0, there are some key challenges that should be taken into account, such as stability of control subsystem, deterministic bounded delay, reliable communication, energy efficiency and limited resources of wireless network \cite{breivold,da2014internet,sheng2015recent}. In control subsystems, the sampled data of sensors is sent to the controller(s), and then the control messages are sent to the actuators from controller(s), through the network. The stability and the performance of the control subsystem can be affected by sampling rate and network based uncertainties like delay, jitter, packet loss, and resource contention  \cite{bello2017guest}.  Therefore, significant researches have focused on determining the sampling rate in control subsystem, guaranteeing deterministic bounded delay and reliability in wireless network  over the past decades \cite{tang2017delay,lu2016real,wang2015delay,yu2017performance,saifullah2014near}.\\
The authors of \cite{saifullah2014near} propose near optimal sampling rates of control plants sharing a limited bandwidth in a WirelesHART network. The authors address the problem of overall control cost minimization while all data flows meet their end-to-end deadlines. In \cite{tang2017delay}, a new routing protocol is proposed to minimize the end-to-end delay for a cognitive radio network that supports real time applications. On the other hand, due to energy crisis, carbon emission concern and also given the fact that most wireless devices relay on batteries, designing new energy-saving network is receiving substantial attention from researchers \cite{sheng2015recent,suto2015energy,kaur2017energy}. However, there is a trade-off between the energy saving and supplying the deterministic bounded delay \cite{chen2017delay}. To guarantee a deterministic delay, a fixed delay bound have to be provided within a connection even in a worst case. A method is proposed in \cite{suto2015energy} to schedule sleep states of wireless computing systems to minimize the power consumption such that a given tolerable delay constraint is not violated. It is assumed in \cite{suto2015energy} that the sampling instants of control subsystem and its tolerable delay are predefined. َAlso, there is another trade-off between the energy saving and the sampling rate. In the case of increasing the sampling rate, the number of sensors sampled data transmissions and control messages will increase and as a result, energy consumption will increase.\par
 Generally, there are three methods to determine the sampling time in a control subsystem: time-triggered (e.g periodic), event-triggered and self-triggered. In event-triggered, instead of periodically updating the sampled data of sensors and control input, a new sample is picked when a threshold, defined by the state of the plant, is violated. On the other hand, under the self-triggered implementations, the next sampling time is computed based on the last state measurement and therefore a constant plant monitoring is not required as in event-triggered method. The self-triggered is more appealing scheme due to its irreplaceable advantage in reducing the number of transmissions between sensors and controller, between controller and actuators and, accordingly, computation load in the controller \cite{heng2015event, heemels2012introduction}.  A new dynamic transmission scheduling is proposed in \cite{tiberi2013energy} to ensure the stability of several processes controlled over a shared IEEE 802.15.4 wireless network and also to reduce the number of sensors sampled data transmissions and increase the sleep time of transmission nodes. A self-triggered sampler is proposed in \cite{tiberi2013energy} to guarantee the stability of each control plant and reduce energy consumption. In addition, in \cite{tiberi2013energy}, it is shown that applying self-triggered sampling may not provide any benefit if the MAC (Media Access Control) layer parameters are not appropriately adjusted. In fact, in contrast to traditional MAC layer designs, the transmission scheduling in CPSs should jointly consider sampling instants, the stability of control subsystem, relevant deterministic delay requirement and network specification \cite{tiberi2013energy,li2014data}.\par
Several standards and relevant MAC protocol have been suggested for wireless communication in Industrial CPSs such as, IEEE 802.15.4, IEEE 802.11ac and WirelessHART so far. However, the fifth generation (5G) may provide an ideal platform for these applications \cite{li2018energy,atat2017enabling}. It is predicted that 5G technologies will be able to support ultra reliable low latency communications (URLLC) and massive machine-type communications (mMTC), which are aligned with industrial wireless network requirements. The 5G technologies may bring considerable benefits to industrial network such as reliability, global connectivity and security. However, a set of challenges need to be addressed before 5G deployment in industrial networks. Two main challenges are scarce number of radio resources and the problem of energy-efficient communication design while providing deterministic delay guarantee \cite{li2018energy,atat2017enabling}. The problem of energy-efficient resource allocation in an industrial wireless network based on 5G technology is considered in \cite{li2018energy}. The deterministic end to end delay requirement is simplified as a minimum transmit rate constraint for each sensor and actuator conservatively. However, this simplification is only applicable if the sampling time in control subsystem is periodic. The number of data transmissions in the case of using periodic sampling is usually more than event triggered and self-triggered, consequently, it results in more power consumption. Determining the sampling time in the control subsystem and allocating the radio resources in the network subsystem, in a energy-efficient manner, have been extensively studied in recent years. However, most of existing works have considered these two issues separately or assumed simple settings \cite{saifullah2014near,suto2015energy,tiberi2013energy,li2018energy}. To the best of our knowledge, the problem of jointly sampling time determining in the control subsystem and power-efficient resource allocation in network subsystem based on 5G technologies, in CPSs, has not been investigated yet. \par
The main purpose of this paper is to minimize consumed power in industrial CPSs. To this aim, we jointly determine the sampling time in control subsystem and allocate the radio resources in network subsystem. We assume an industrial CPS which is composed of several control plants and Rate Constrained (RC) users that share a single cell Orthogonal Frequency-Division Multiple Access (OFDMA) network. The major contributions of this paper are as follows: 
 \begin{itemize} 
 	
 	\item The problem of jointly determining of sampling time in control subsystem and allocating radio resources in network subsystem, in the CPS, is formulated. The problem is a multi objective optimization with aims of determining the next maximum allowable sampling time of control plants and minimizing the aggregate power consumption in uplink and downlink, considering  the dynamics and desired performance of each control plant, the quality of service requirement of RC users, power and sub-carrier constraints. The general goal of the problem is power consumption minimization through decreasing sampling rate and power-efficient resource allocation. 
 	
 	\item To solve the multi objective optimization problem, a novel approach is proposed. The proposed approach decomposes the optimization problem into two smaller loosely coupled problems. One of the two decomposed problems is determining the next maximum allowable sampling time of control plants and another one is the resource allocation problem. The approach is launched from solving the control subsystem problem using the self-triggered method. Based on using the self-triggered method, power saving is improved by minimizing the number of sampling time and switching some sensors and actuators to sleep mode when they are not working and wake them up when needed. The approach is continued to solve power and sub-carrier allocation problem in the OFDMA wireless network with the goal of power consumption minimization on the condition of the deterministic delay which guarantees stability and desired  performance of control subsystem. 
 	
 	\item The power and sub-carrier allocation problem is formulated with aims of minimizing aggregate consumed power (transmit and circuit power) in uplink and downlink considering QoS (Quality of Service) requirement of RC users and deterministic end to end delay requirement for each control plant. The QoS requirement is considered in term of a minimum transmit rate. Thanks to self-triggered method, the deterministic end to end delay requirement guarantees the stability and desired performance of control subsystem. The problem with aims of power consumption minimization in uplink and downlink is also a multi objective optimization. The weighted sum technique is applied to convert two objective functions into a single objective function. The resultant single objective problem is mixed-integer nonlinear programing (MINLP). Directly solving the MINLP problems suffers from high computational complexity. To deal with it, two steps iterative approach is used which sequentially performs power allocation and sub-carrier allocation in each iteration. These two steps are proceeded iteratively until the predetermined convergence criterion is satisfied.
 	
 	\item We show the power saving effectiveness of using the proposed approach via simulation results. Simulation results demonstrate that our approach significantly improves total power saving in the CPS, while it is shown that all control plants are properly controlled. 
 	
 \end{itemize}
\textbf{\textit{Notations:}}
"$||x||$" stands for Euclidean norm of the vector $x$. $\dot{x}(t)$ represents the derivative of variable \textit{x(t)} with respect to the original time, \textit{t}.\\ 
\indent The organization of the paper is as follows: The system model is described in section II. The problem is formulated in Section III. In Section IV, the proposed approach is presented. The proposed approach is evaluated through simulation results in Section V. Finally, the paper is concluded in Section VI.\\

\vspace{10pt}
\section{SYSTEM MODEL} 
\vspace{10pt}
 Consider a model of CPSs comprising 1) $I$ numbers of independent plants, 2) an OFDMA cellular network and 3) a central controller (Fig.1). Each plant $i$, $i\in \{1,2,...,I\}$, has multiple sensors and multiple actuators that communicate with the controller via  Remote Transmitter Units (RTUs) through the OFDMA network (Fig.1). RTUs of each plant are connected to the corresponding sensors and actuators via a wired network in which the time delay between the RTUs and the relative sensors and actuators is ignorable. Given plant $i$, let $y_{i,k}$ be the $k^{th}$ sampled data measured by sensors of plant $i$ which is sent to the controller at sampling time of $t_{i,k}$ via the corresponding RTU, through the OFDMA network.  
 The controller calculates an appropriate control input ($u_{i,k}$) based on the measurements ($y_{i,k}=y_i(t_{i,k})$). The controller sends $u_{i,k}$ to the relevant RTU (which is wired to actuators of plant $i$) through the OFDMA network. We assume that the OFDMA network is responsible for all data exchanges in the system model, including the communication between RTUs and the controller as well as the communication between other users which send RC or Best Effort (BE) packets.\\
   Based on physical infrastructure of ITU's (International Telecommunication Union) 5G architecture, RAN (Radio Access Network) real time functions including computing and data storing must be deployed on the central office Data Center layer, located close to the base station and users \cite{5GArchi}. So we assume the central controller run on a dedicated platform or general platform in the central office Data Center layer.\\
    This system model is very appropriate for plenty of emerging industrial cyber physical systems such as smart industry automation and smart grid. In the following subsections, the CPS model are explained in details.
 \begin{figure}[h!]
 	\centerline{\includegraphics[width=0.5\textwidth]{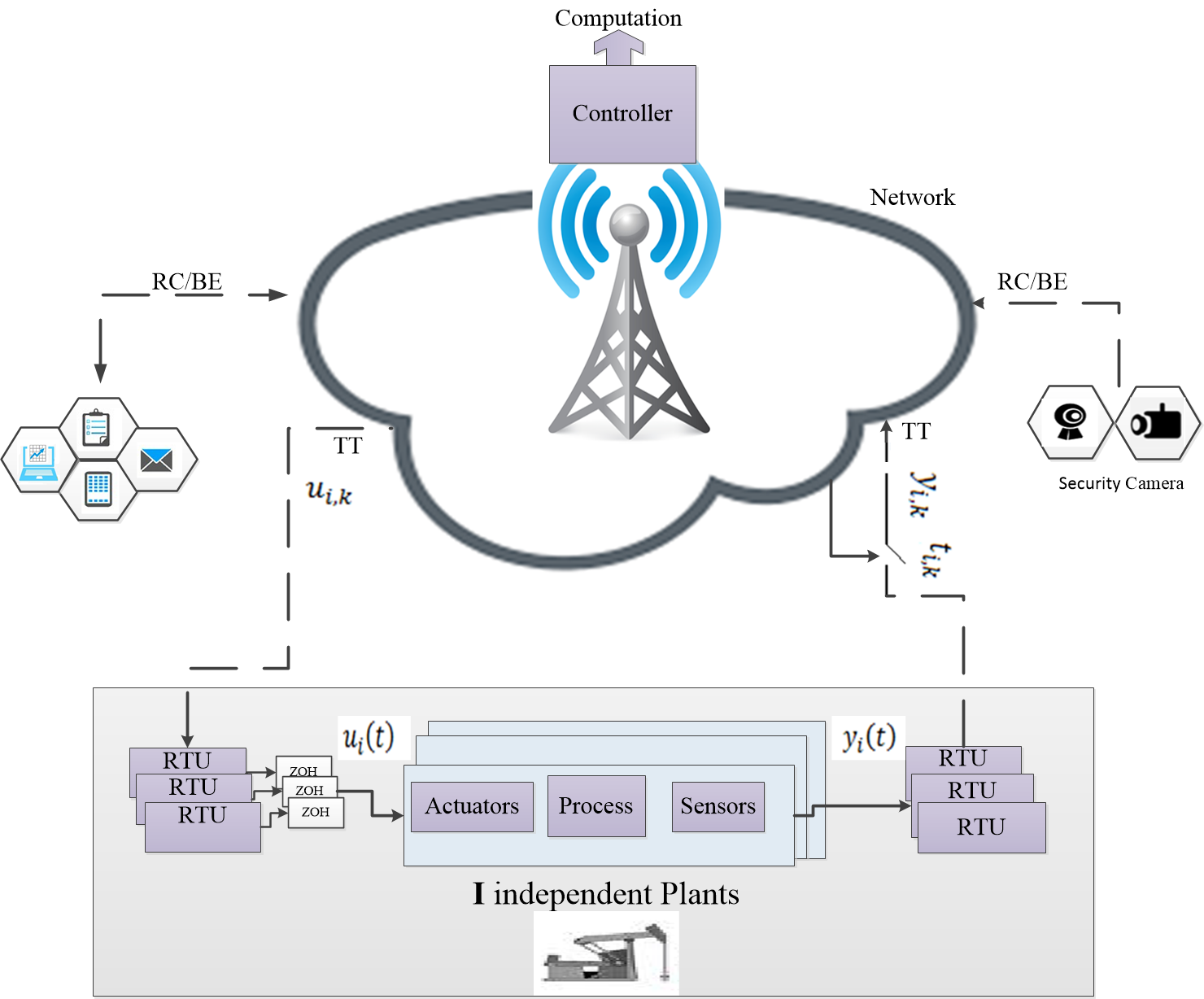}}
 	\label{fig:System model}
 	\caption{System model of the considered CPS}
 \end{figure}
\subsection{Plants}
 Each plant $i$ has $n_i$ states, $m_i$ actuators and $q_i$ measurable states ($q_i\leq n_i$) which measured by $q_i$ sensors. The dynamics of each plant $i$ is linear in the form of:
 \begin{equation}
 \begin{split}
 \dot{x} & = A_i x_i(t) + B_i u_i(t) + d_i(t),\\
  y_i(t) & = C_i x_i(t),
 \label{eqn:pl1}
 \end{split}
 \end{equation}	 
where $x_i(t)\in \R^{n_i}$, $u_i(t)\in \R^{m_i}$ and $y_i(t)\in \R^{q_i}$ are the vector of the plant states, the control input and plant output (measured by sensors) at time $t\in \R^+ $, respectively. $d_i(t)\in \R^{m_i}$ is used to model measurement and actuation disturbances and other sources of uncertainty. $A_i\in \R^{n_i\times n_i}$, $B_i\in \R^{n_i \times m_i}$ and $C_i\in \R^{q_i \times n_i}$ are state, control and output matrix, respectively. We suppose that the disturbance is non measurable and bounded $\norm{d_i(t)}\leq d$. 
 \subsection{Controller}    
 The central controller receives measurements ($y_{i,k}$), sent by sensors, and uses them to calculate the control input ($u_{i,k}$), afterward sends the control input to the relevant RTU. 
  The control input is calculated according to the following control law:
 \begin{equation}
  u_{i,k} = k_ix_i (t=t_{i,k}),
 \label{eqn:conlsamp1}
 \end{equation} 
 where $k_i$ is calculated by the controller for plant $i$ such that the dynamics of plant $i$ is stable ($||x(t)||<\infty: \forall t $).
  The control input used by actuators is held constant by ZOH (Zero Order Hold) between two successive control input updates. Therefore the control input used by actuators is a piecewise constant signal:
   \begin{equation}
   u_i(t) = u_{i,k},\quad t \in [t_{i,k}+\Delta^{total}_{i,k},\quad t_{i,k+1}+\Delta^{total}_{i,k+1}),\\
   \label{eqn:conlsamp}
   \end{equation}  
 where $\Delta^{total}_{i,k}$ denotes the total delay (end to end delay) for plant $i$ in the $k^{th}$ sampling.
  Considering (\ref{eqn:conlsamp1}) and (\ref{eqn:conlsamp}), (\ref{eqn:pl1}) may be rewritten as bellow for $ t \in [t_{i,k}+\Delta^{total}_{i,k},t_{i,k+1}+\Delta^{total}_{i,k+1}) $:  
   \begin{equation}
    \begin{split}
    \dot{x_i} & = A_i x_i(t) + B_i u_{i,k} + d_i(t),\\
    \dot{x_i} & = (A_i+B_i) x_i(t) + B_ik_ie_{i,k}(t) + d_i(t),\\
    e_{i,k}(t) & = x_i(t)-x_i(t=t_{i,k}), \quad t \in [t_{i,k}+\Delta^{total}_{i,k},t_{i,k+1}+\Delta^{total}_{i,k+1}),
    \label{eqn:pl2}
    \end{split}
    \end{equation}	 
    where $e_{i,k}(t)$ denotes the deviation between $x_i(t)$ and $x_i(t=t_{i,k})$ during $ t \in [t_{i,k}+\Delta^{total}_{i,k},t_{i,k+1}+\Delta^{total}_{i,k+1}) $ which is considered as an index of plant performance. $e_{i,k}(t)$ is bounded by $\delta_i$, that is: 
    \begin{equation}
    ||e_{i,k}(t)||\leqslant\delta_i,
    \label{eqn:bound}
    \end{equation}
     where $\delta_i$ denotes the maximum deviation of  $x_i(t)$ from $x_i(t=t_{i,k})$ during $t\in [t_{i,k}+\Delta^{total}_{i,k},t_{i,k+1}+\Delta^{total}_{i,k+1})$ which is tolerated by plant $i$.\\   
 \subsection{Network}
 Consider a single cell OFDMA network which contains two types of users namely, RC users and the control system users (Time Constrained-TC). RC users send Rate Constrained or Best Effort packets and are irrelevant to the control system. It is assumed that the network consists of \textit{M} uplink users and \textit{N} users in downlink, forming the sets $\mathcal{U}_u=\{1,2,...,M\}$ and $\mathcal{U}_d=\{1,2,...,N\}$, respectively. Uplink users include a set of $M^c$ RTUs (control system users) denoted by $\mathcal{U}^c_u=\{1,2,...,M^c\}$ and a set of $M^u$ RC users denoted by $\mathcal{U}^u_u=\{1,2,...,M^u\}$. Users in downlink include a set of $N^c$ RTUs (control system users) denoted by $\mathcal{U}^c_d=\{1,2,...,N^c\}$ and a set of $N^u$ RC users denoted by $\mathcal{U}^u_d=\{1,2,...,N^u\}$. It is assumed that each plant enjoys two RTUs, one for sending its sensors data in uplink and another one for receiving its control input data in downlink, hence $M^c=I$ and $N^c=I$. The total bandwidth is divided in $L$ sub-carriers in set  $\mathcal{L}=\{1,2,...,L\}$. The channel gain for the uplink user $m$ and the downlink user $n$ on the sub-carrier $l$ are represented by $g_{m,l}$ and $g_{n,l}$, respectively. Let $p^u_{m,l}$ be the transmit power of the uplink user $m$ on the sub-carrier $l$ and $p^d_{n,l}$ be the BS transmit power for the downlink user $n$ on the sub-carrier $l$, consecutively. In addition $\overline{P}_m$ and $\overline{P}_{BS}$ denote the peak transmit power of the downlink user $m$ and the BS, respectively ($\forall m:\sum_{l=1}^{L}p^u_{m,l}\leq\overline{P}_m$, and $\sum_{n=1}^{N}\sum_{l=1}^{L}p^d_{n,l}\leq\overline{P}_{BS}$). 
 
 The uplink transmission rate for the uplink user $m$ is given by:
 \begin{equation}
 	R^u_m=\sum_{l=1}^{L}\alpha_{m,l}wlog(1+\frac{p^u_{m,l}g_{m,l}}{N_0}),
 	\label{eqn:rateu}
  \end{equation}
  where $\alpha_{m,l}$ is a binary variable and $\alpha_{m,l}=1$ if sub-carrier $l$ is allocated to the uplink user $m$; otherwise, $\alpha_{m,l}=0$. Additionally $N_0$ is the power of additive white Gausian noise and $w$ denotes the bandwidth of each sub-carrier.  
 The downlink transmission rate for the downlink user $n$ is given by:
 \begin{equation}
  R^d_n=\sum_{l=1}^{L}\beta_{n,l}wlog(1+\frac{p^d_{n,l}g_{n,l}}{N_0}),
 	\label{eqn:rated}
 \end{equation}
 where $\beta_{n,l}$ is a binary variable and $\beta_{n,l}=1$ if sub-carrier $l$ is assigned to the downlink user $n$; otherwise, $\beta_{n,l}=0$.\\ 
 The QoS requirement of RC users is stated in term of a minimum transmit rate. The required QoS of RC users in uplink and downlink is given by:
 \begin{equation}
 \begin{aligned}
 R^u_m &\geq\overline{R}^u_{m},\forall m\in \mathcal {U}^u_u,\\
 R^d_n &\geq\overline{R}^d_{n},\forall n\in \mathcal {U}^u_d.
\end{aligned}
 \label{eqn:ratecons}
\end{equation}
  \indent Given $k^{th} $ sampling of plant $i$, we consider three types of delays in this system model including uplink transmission delay, downlink transmission delay and computation delay, denoted by $\Delta^u_{i,k}$, $\Delta^d_{i,k}$ and $\Delta^{comp}_{i,k}$, respectively. Let $\Delta^{comp}_{max}$ be the maximum computation delay that is $\Delta^{comp}_{i,k}\leq\Delta^{comp}_{max}$. Conservatively, we assume $\Delta^{total}_{i,k}$ is given by:
 \begin{equation}
 \Delta^{total}_{i,k}=\Delta^u_{i,k}+\Delta^d_{i,k}+\Delta^{comp}_{max}.
 \label{eqn:delay}
 \end{equation}
  
 The RC and TC users send and receive data with size of $L_{RC}$ and $L_{TC}$ bit, respectively. At each $k^{th} $ sampling of plant $i$, uplink and downlink transmission delay are given by:
  \begin{equation}
   \begin{aligned}
  \Delta^u_{i,k}&=\frac{L_{TC}}{R^u_i}, \\ 
  \Delta^d_{i,k}&=\frac{L_{TC}}{R^d_i},
  \end{aligned}
  \label{eqn:up-delay}
  \end{equation}
 where $R^u_i$ and $R^d_i$ are the uplink and downlink transmission rate for the user of plant $i$ given by (6) and (7), respectively.
 It is assumed that uplink users' total power consumption of the CPS may be obtained in terms of uplink power matrix ($\textbf{P}^u{\mathop:}=[{p^u_{m,l}}_{(m\in U_u,l\in \textit{L})}]$) and uplink sub-carrier assignment matrix ($\textbf{A}^u{\mathop:}=[{\alpha_{m,l}}_{(m\in U_u,l\in \textit{L})}]$) as the following:  
 \begin{equation}
 P^u_{total}(\textbf{A}^u,\textbf{P}^u)=\sum_{m=1}^{M}P^{cst}_m+\sum_{m=1}^{M}\sum_{l=1}^{L}\alpha_{m,l}p^u_{m,l},
 	\label{eqn:poweU}
  \end{equation}
 where $P^{cst}_m$ is the constant circuit power consumed by uplink user $m$. Also the BS's total power consumption at the modeled CPS is given by:
 \begin{equation}
 P^{BS}_{total}(\textbf{B}^d,\textbf{P}^d)=P^{cst}_{BS}+\sum_{n=1}^{N}\sum_{l=1}^{L}\beta_{n,l}p^d_{n,l},
 \label{eqn:poweD}
 \end{equation}
 where $P^{cst}_{BS}$ is the constant circuit power consumed by BS, ($\textbf{P}^d{\mathop:}=[{p^d_{n,l}}_{(n\in U_d,l\in \textit{L})}]$) and ($\textbf{B}^d{\mathop:}=[{\beta_{n,l}}_{(n\in U_d,l\in \textit{L})}]$) are downlink power matrix and downlink sub-carrier assignment matrix, respectively.\\
 
 \section{Problem Formulation}
 We consider the problem of consumed power minimization in the CPS. To realize consumed power minimization in the CPS, the problem of jointly decreasing the number of sampling time in control subsystem and allocating the radio resources in network subsystem, in a power-efficient manner, is considered. The objectives are determining the next maximum allowable sampling instant of each plant, $t_{i,k}$, and minimizing aggregate power consumption in uplink and downlink subject to the dynamics and desired performance requirement of each plant, QoS of RC users, power and sub-carrier constraints. The multi objective optimization problem is formulated as follows:\\
 \begin{equation}
 \begin{aligned}
 &{\text{maximize }}t_{i,k} \quad :\forall i\in \{1,2,...,I\}\\
 &\underset{\textbf{A}^u,\textbf{P}^u}{\text{minimize }}P^u_{total}\\
 &\underset{\textbf{B}^d,\textbf{P}^d}{\text{minimize }}P^{BS}_{total}\\
 &\text{subject to}:\\
 &C_{1}:\dot{x_i}= (A_i+B_i) x_i(t) + B_ik_ie_{i,k}(t) + d_i(t)\\ 
 & \quad : t \in [t_{i,k-1}+\Delta^{total}_{i,k-1},t_{i,k}+\Delta^{total}_{i,k})\\
 &C_{2}:||e_{i,k}(t)||\leqslant\delta_i\\
 &C_{3}:\sum_{m=1}^{M}\alpha_{m,l}\leq 1,  \forall l\in \mathcal {L}\\
 &C_{4}:\sum_{n=1}^{N}\beta_{n,l}\leq 1, \forall l\in \mathcal {L}\\
 &C_{5}:\alpha_{i,k}\in\{0,1\},\forall i,k\\
 &C_{6}:\beta_{i,k}\in\{0,1\},\forall i,k\\
 &C_{7}:\sum_{l=1}^{L}p^u_{m,l}\alpha_{m,l}\leq\overline{P}_m,\forall m\in \mathcal {U}_u\\
 &C_{8}:\sum_{n=1}^{N}\sum_{l=1}^{L}p^d_{n,l}\beta_{n,l}\leq\overline{P}_{BS}\\
 &C_{9}:\frac{1}{2}R^u_m\geq\overline{R}^u_{m},\forall m\in \mathcal {U}^u_u\\
 &C_{10}:\frac{1}{2}R^d_n\geq\overline{R}^d_{n},\forall n\in \mathcal {U}^u_d,
 \label{eqn:opt}
 \end{aligned}
 \end{equation}
 where the first two constraints denote the dynamics and the desired performance requirement for each plant, respectively. The constraints $C_{3}$, $C_{4}$, $C_{5}$ and $C_{6}$ correspond to the exclusive sub-carrier allocation in uplink and downlink, respectively. It is noticeable that the same subset of sub-carriers is used in uplink and downlink during the distinct time intervals. The constraints $C_{7}$ and $C_{8}$ relate to the maximum tolerated power consumption in each uplink user and BS. The last two constraints guarantee the required QoS of RC users in uplink and downlink consequently. Due to the similarity of uplink and downlink time length, we share the total time length of transmission between uplink and downlink equally. The factor of $\frac{1}{2}$ in $C_{7}$ and $C_{8}$ denotes this sharing.\\
 The optimization problem in (\ref{eqn:opt}) is a multi objective mixed integer nonlinear programing. To address the multi objective optimization, we decompose the original optimization problem into two smaller loosely coupled problems, as is explained in the following section.
  \vspace{10pt}
  \section{The Proposed Approach}
  \vspace{10pt}
 We decompose the optimization problem into two smaller loosely coupled problems. One of the two decomposed problems is determining the next maximum allowable sampling time of control plants and another one is the power-efficient resource allocation problem (Fig.2). First, the control subsystem problem is addressed by using the self-triggered method. We use self-triggered method to reduce the transmission number of control subsystem nodes and increase their sleeping time. By determining the maximum allowable sampling instant ($t_{i,k}$), it is possible to put some nodes in sleep mode when they are not needed. Then, having determined the number of registered plants to communicate at $k^{th}$ sampling time and their relevant end to end delay requirement, the problem of resource allocation is addressed. This problem is a joint power and sub-carrier allocation in uplink and downlink with aims of uplink and downlink power minimization, considering the QoS requirement of RC users and the deterministic end to end delay requirement for the plants which was determined in previous step ($\Delta^i_{max}$). Thanks to self-triggered method, the deterministic delay requirement guarantees stability and desired  performance of the plants.\\
 It is possible to enlarge feasible region by increase in $\delta_i$ parameter such that $\delta_i \leq \delta_{i,max} $. By increasing $\delta_i$, the inter-sampling time of the plant $i$ will increase and the number of plants wanting to communicate will decrease. $R^u_m$ and $R^d_n$ are determined through the solution of the resource allocation problem and so $\Delta^{total}_{i,k}$ is revealed. $\Delta^{total}_{i,k}$ will be used to determine the next sampling instant $t_{i,k+1}$.
   \begin{figure}[h!]
 	\centerline{\includegraphics[width=0.5\textwidth]{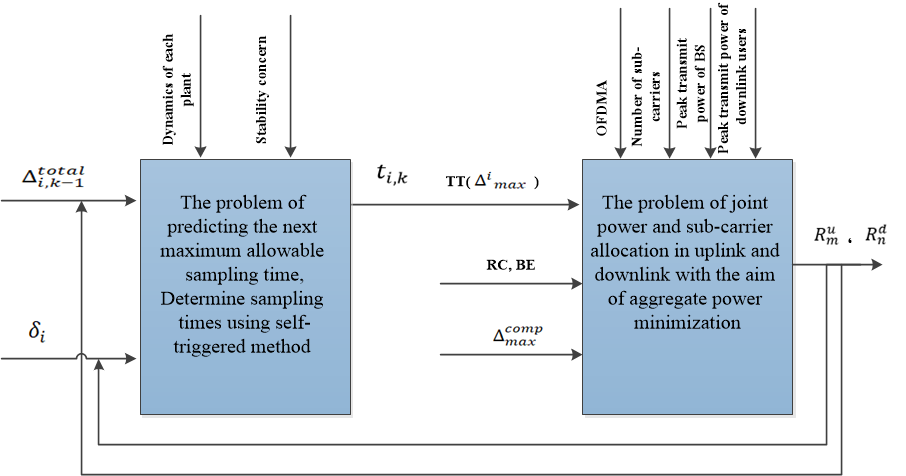}}
 	\label{fig:Proposed Approach}
 	\caption{Proposed approach}
 \end{figure} 
 
 \subsection{The Sampling Time Determining Problem}
 In the problem of determining the sampling time, we aim to determine the next maximum allowable sampling time of control plants. The problem is formulated as:
 \begin{equation}
 \begin{aligned}
 &{\text{maximize }}t_{i,k} \quad :\forall i\in \{1,2,...,I\}\\
 &\text{subject to}:\\
 &C_{1}:\dot{x_i}=(A_i+B_i) x_i(t) + B_ik_ie_{i,k}(t) + d_i(t)\\ 
 & \quad : t \in [t_{i,k-1}+\Delta^{total}_{i,k-1},t_{i,k}+\Delta^{total}_{i,k}),\\
 &C_{2}:||e_{i,k}(t)||\leqslant\delta_i.\\
 \label{eqn:predict}
 \end{aligned}
 \end{equation}
 The self-triggered method is used to solve this problem \cite{heng2015event}. We use the proposed self-triggered method in \cite{tiberi2013energy}. According to self-triggered method, the next maximum allowable sampling time is determined using the plant dynamics model, its current state ($x_i(t=t_{i,k})$) and the total delay (end to end delay) in the previous sampling time while the stability and desired performance of the plant is guaranteed \cite{tiberi2013energy}. As stated, the proposed self-triggered sampling method in \cite{tiberi2013energy} is adopted:
 \begin{equation}
 \begin{aligned}
  t_{i,k+1} =&t_{i,k} + min(\gamma(x_{i,k-1},x_{i,k},\hat{d}_{i,k-1},\hat{d}_{i,k},\\
 &\Delta^{total}_{i,k}),h_{max}),
  \label{eqn:sampltime}
 \end{aligned}
 \end{equation}
 where $\hat{d}_{i,k}$ and $\hat{d}_{i,k-1}$ denote the estimation of $d_i(t=t_k)$ and $d_i(t=t_{k-1})$ respectively, and $h_{max}$ is the maximum acceptable interval between two consecutive sampling times for plant $i$ by which the stability of plant $i$ and $||e_{i,k}(t)||\leqslant\delta_i$ are satisfied. $ \gamma(x_{i,k-1},x_{i,k},\hat{d}_{i,k-1},\hat{d}_{i,k},\Delta^{total}_{i,k})$ is defined as follows: 
 \begin{equation}
 \begin{aligned}
 \gamma(x&_{i,k-1},x_{i,k},\hat{d}_{i,k-1},\hat{d}_{i,k},\Delta^{total}_{i,k})\\
 \mathrel{\mathop:}=&\frac{1}{||A||}\ln \left(\frac{\phi (x_{i,k},\hat{d}_{i,k})}{\upsilon(x_{i,k-1},x_{i,k},\hat{d}_{i,k-1},\hat{d}_{i,k},\Delta^{total}_{i,k})}\right)\\
 +&\Delta^{total}_{i,k}-\Delta^i_{max},
 \label{eqn:landa2}
 \end{aligned}
 \end{equation}
  where $\Delta^i_{max}$ is the maximum total delay (delay between sending instant of measurements and receiving instant of input by actuators) tolerated by plant $i$ for all $k$. $\phi(x_{i,k-1},x_{i,k},\hat{d}_{i,k-1},\hat{d}_{i,k},\Delta^{total}_{i,k})\mathrel{\mathop:}=||A_i||\delta_i+||(A_i+B_ik_{i,k})x_{i,k}||+||\hat{d}_{i,k}||$ and $\upsilon$ is defined as follows:
  \begin{equation}
 \begin{aligned}
 \upsilon(x_{i,k-1}&,x_{i,k},\hat{d}_{i,k-1},\hat{d}_{i,k},\Delta^{total}_{i,k})\\   \mathrel{\mathop:}=&||A_ix_{i,k}-B_ik_{i,k}x_{k-1}||+||\hat{d}_{i,k}||*\\
   &(\exp(||A_i||\Delta^{total}_{i,k})-1)||(A_i+B_ik_{i,k})x_{i,k}||\\
 +&||\hat{d}_{i,k}||.
  \label{eqn:landa1}
  \end{aligned}
   \end{equation}
If the controller does not receive any new update from plant $i$ after $h_{max}$ seconds, it will notify the relevant RTU to send the update.\\
 In summing up, the controller uses (\ref{eqn:sampltime}) to determine the time of sampling $t_{i,k}$. The stability of plant $i$ and the desired performance are satisfied if $\Delta^{total}_{i,k}\leq\Delta^i_{max}$. The controller awakes the RTU wired to sensors of plant $i$, at  time of $t_{i,k}$, to pick $k^{th}$ measurement of plant $i$ ($y_{i,k}=y_i(t_{i,k})$) and calculates an appropriate control input ($u_{i,k}$) based on the measurements ($y_{i,k}=y_i(t_{i,k})$). The controller sends $u_{i,k}$ to the relevant RTU (which wired to plant $i$ actuators) through the OFDMA network.
 
 \subsection{Power-Efficient Resource Allocation Problem}
 The problem of power-efficient resource allocation is formulated as:
  \begin{equation}
 \begin{aligned}
  &\underset{\textbf{A}^u,\textbf{P}^u}{\text{minimize }}P^u_{total}\\
 &\underset{\textbf{B}^d,\textbf{P}^d}{\text{minimize }}P^{BS}_{total}\\
 &\text{subject to}:\\
 &C_{3}:\sum_{m=1}^{M}\alpha_{m,l}\leq 1,  \forall l\in \mathcal {L}\\
 &C_{4}:\sum_{n=1}^{N}\beta_{n,l}\leq 1, \forall l\in \mathcal {L}\\
 &C_{5}:\alpha_{i,k}\in\{0,1\},\forall i,k\\
 &C_{6}:\beta_{i,k}\in\{0,1\},\forall i,k\\
 &C_{7}:\sum_{l=1}^{L}p^u_{m,l}\alpha_{m,l}\leq\overline{P}_m,\forall m\in \mathcal {U}_u\\
 &C_{8}:\sum_{n=1}^{N}\sum_{l=1}^{L}p^d_{n,l}\beta_{n,l}\leq\overline{P}_{BS}\\
 &C_{9}:\frac{1}{2}R^u_m\geq\overline{R}^u_{m},\forall m\in \mathcal {U}^u_u\\
 &C_{10}:\frac{1}{2}R^d_n\geq\overline{R}^d_{n},\forall n\in \mathcal {U}^u_d,\\
 &C_{11}:\Delta^{comp}_{max}+\frac{L_{TC}}{\frac{1}{2}R^u_m}+\frac{L_{TC}}{\frac{1}{2}R^d_n}\leq\Delta^i_{max},\\
 &\forall m\in  \mathcal {U}^c_u , \forall n\in  \mathcal {U}^c_d, \forall i \in  \mathcal {I}^r
 \label{eqn:optresource}
 \end{aligned}
 \end{equation}
 where $C_{11}$ is the deterministic end to end delay requirement for the set of plants ($\mathcal I^r$) determined by using (\ref{eqn:sampltime}) to solve (\ref{eqn:predict}). $\mathcal I^r$ denotes the set of plants that their RTUs should communicate with controller in the time of solving (\ref{eqn:optresource}). Thanks to self-triggered method, the stability and desired performance of the plants are ensured provided that the $C_{11}$ constraint is satisfied. As stated before, the factor of $\frac{1}{2}$ in $C_{9}$, $C_{10}$ and $C_{11}$ denotes the equal time length sharing between uplink and downlink.\\
 We use weighted sum technique to transform two objective functions into a single objective function \cite{cho2017survey}. Therefore the transformed objective function can be written as:
 \begin{equation}
 \begin{aligned}
 &\underset{\textbf{A}^u,\textbf{B}^d,\textbf{P}^u,\textbf{P}^d}{\text{minimize}}aP^u_{total}+(1-a)P^{BS}_{total}\\
 &\text{subject to}:C_{3}-C_{11},
 \label{eqn:opt2}
 \end{aligned}
 \end{equation}
 where $0\leq a\leq 1$ and denotes the priority of objective functions. 
 The optimization problem in (\ref{eqn:opt2}) is non-convex due to coupled continuous and integer variable in both the objective function as well as the constraints. Notice that (\ref{eqn:opt2}) is a Mixed Integer Nonlinear Programming (MINLP) which can be solved using exhaustive search over all sub-carrer assignment choices. However the complexity of MINLP problems raises fast when the number of users and sub-carriers increases. In order to tackle the computational complexity of (\ref{eqn:opt2}), we transform the problem into two subproblems: i) sub-carrier assignment and ii) power allocation. In other words, for a given power allocation, elements of sub-carrier assignment matrices ($\textbf{A}^u$ and $\textbf{B}^d$) are considered as variables in the first step. Thereafter, the power matrices of users and BS ($\textbf{P}^u$ and $\textbf{P}^d$) are assumed variables while using sub-carrier assignments given in previous step. The two steps are proceeded iteratively until convergence is attained \cite{venturino2009coordinated, mokari2016limited}. This approach has been widely used for resource allocation problems \cite{venturino2009coordinated, mokari2016limited, son2011refim, tweed2017outage}.\\
 
 \textbf{Sub-carrier Assignment Subproblem:} The sub-carrier assignment subproblem for a given power allocation can be written as follows:
   \begin{equation}
  \begin{aligned}
  &\underset{\textbf{A}^u,\textbf{B}^d}{\text{minimize }} aP^u_{total}+(1-a)P^{BS}_{total}\\
  &\text{subject to}: C_{3}-C_{11}.\\
  \label{eqn:opt3}
  \end{aligned}
  \end{equation}
The problem of (\ref{eqn:opt3}) is integer nonlinear programming. We solve this problem using MATLAB/CVX with MOSEK solver \cite{ANU:2019}.\\ 

\textbf{Power Allocation subproblem:} Given a sub-carrier assigned in (\ref{eqn:opt3}), the power allocation subproblem is stated as:
\begin{equation}
\begin{aligned}
&\underset{\textbf{P}^u,\textbf{P}^d}{\text{minimize }} aP^u_{total}+(1-a)P^{BS}_{total}\\
&\text{subject to}: C_{7}-C_{11}.\\
\label{eqn:opt4}
\end{aligned}
\end{equation}
 The objective function is linear with respect to the transmit power variables. This subproblem is convex optimization problem and MATLAB/CVX can be applied to numerically solve the optimal problem.
 
 \section{Simulation Result and Discussion} 
The efficacy of the proposed approach is evaluated through numerical simulations. We consider three independent plants that communicate over an OFDMA cellular network and the simulation time of about 50s for each simulation. The control plants and network specification are explained below:\\

\textbf{OFDMA cellular network:} We consider a single cell network where 5 RC users, 6 sensors and 6 actuators are randomly spread within the square cell. The channel gain for each uplink (downlink) user is modeled as $g_{m,l}= \overline{d}_{m,l}^{-3}\overline{h}$ ($g_{n,l}= \overline{d}_{n,l}^{-3}\overline{h}$), where $\overline{d}_{m,l}$ ($\overline{d}_{n,l}$) denote the distance between the transmitter and receiver, $\overline{h}$ is the attenuation factor that indicates power variations and $\overline{h}=0.09$. The simulation parameters of the considered network are listed in Table I.\\

\begin{table}\begin{center}
		\caption{Values of Network parameters.}
		\begin{tabular}{||c  c ||} 
			\hline
			Parameter&Value \\ [0.7ex] 
			\hline\hline
			Bandwidth of each sub carrier & 180KHz \\ 
			\hline
			Maximum transmit power of each user ($\overline{P}_m$) & 23dBm \\
			\hline
			Maximum transmit power of BS ($\overline{P}_{BS}$) & 43dBm\\
			\hline
			Constant Power of each user ($P^{cst}_m$) & 0.1dBm \\
			\hline
			Constant Power of BS ($P^{cst}_{BS}$) & 20dBm  \\
			\hline
			Noise Power ($N_0$) & -62.24dBm \\
			\hline
			Distance between any user to BS & 10-50m \\
			\hline
			Minimum data rate for RC users in uplink ($\overline{R}^u_{m}$) & 50 bit/s \\
			\hline
			Minimum data rate for RC users in downlink ($\overline{R}^d_{n}$) & 100 bit/s \\
			\hline
			Maximum  total delay ($\Delta^i_{max}$) & 1s \\
			\hline
			Length of TC users ($L_{TC}$) & 70 bit \\
			\hline
			Number of sub-carries (L) & 16\\ [1ex] 
			\hline
		\end{tabular}
		\label{tab: Table1}
\end{center}\end{table} 

\textbf{Control Plants:} Similar to \cite{tiberi2013energy}, we consider three control plants as bellow:
\subsubsection{First Control plant} 
 The state and control matrix of first control plant are set as bellow:
 \begin{equation}
 \begin{aligned}
  A_1=
 \begin{bmatrix}
 -0.1 & 0.05 \\
 0.2 & 0.1 
 \end{bmatrix},
 B_1=
 \begin{bmatrix}
 0 \\
 1  
 \end{bmatrix}\\ 
  \end{aligned}
 \label{eqn:matix1}
 \end{equation}
 The eigenvalues of first closed control loop are placed in $\lambda_1^1(A_1+B_1K_1)=-0.25$, $\lambda_2^1(A_1+B_1K_1)=-0.18$. The upper bound of external disturbance is considered 0.6 and the initial values are set $x_1^1(0)=-20$, $x_2^1(0)=15$.\\
 \subsubsection{Second Control plant} 
 The state and control matrix of second control plant are set as bellow:
 \begin{equation}
 \begin{aligned}
 A_2=
 \begin{bmatrix}
 0.01 & 0.2 \\
 0.03 & 0 
 \end{bmatrix},
 B_2=
 \begin{bmatrix}
 1 \\
 1  
 \end{bmatrix}\\ 
 \end{aligned}
 \label{eqn:matix2}
 \end{equation}
 The eigenvalues of second closed control loop are placed in $\lambda_1^2(A_2+B_2K_2)=-0.15$, $\lambda_2^2(A_2+B_2K_2)=-0.3$. The upper bound of external disturbance is considered 1.2 and the initial values are set $x_1^2(0)=-12$, $x_2^2(0)=12$.\\
 \subsubsection{Third Control plant} 
 The state and control matrix of third control plant are set as bellow:
 \begin{equation}
 \begin{aligned}
 A_3=
 \begin{bmatrix}
 0.2 & 0.01 \\
 0.3 & -0.8 
 \end{bmatrix},
 B_2=
 \begin{bmatrix}
 1 \\
 2  
 \end{bmatrix}\\ 
 \end{aligned}
 \label{eqn:matix3}
 \end{equation}
 The eigenvalues of third closed control loop are placed in $\lambda_1^3(A_3+B_3K_3)=-0.4$, $\lambda_2^3(A_2+B_2K_2)=-0.6$. The upper bound of external disturbance is considered 0.55 and the initial values are set $x_1^3(0)=-5$, $x_2^3(0)=4$.\\
 Each plant has two sensors and two actuators. We assume one central controller in the central office layer which is located close to BS.\\ 
 First, in table II, we compare the number of sensors transmission of each plant when the proposed approach (using self-triggered method) and periodic method are applied for a simulation time of 50s.
 \begin{table}\begin{center}
 		\caption{The number of sensors transmission of each plant when the proposed approach and periodic method are user.}
 		\begin{tabular}{|c c c|} 
 			\hline
 			Plant&Proposed Approach &Periodic Method\\ [0.5ex] 
 			\hline
 			First Plant &37 &555\\ 
 			\hline
 			Second Plant &31 &555\\
 			\hline
 			Third Plant &25 &555\\[0.5ex]
 			\hline
 		\end{tabular}
 		\label{tab: Table2}
 \end{center}\end{table}   
We compare responses of the CPS when proposed approach and periodic method  are used in Fig.3. The results demonstrate that all control plants are properly controlled. Although the number of transmissions is reduced considerably in proposed approach, the control plants responses are similar to periodic method implementation.
 \begin{figure}[h!]
 	\centerline{\includegraphics[width=0.5\textwidth]{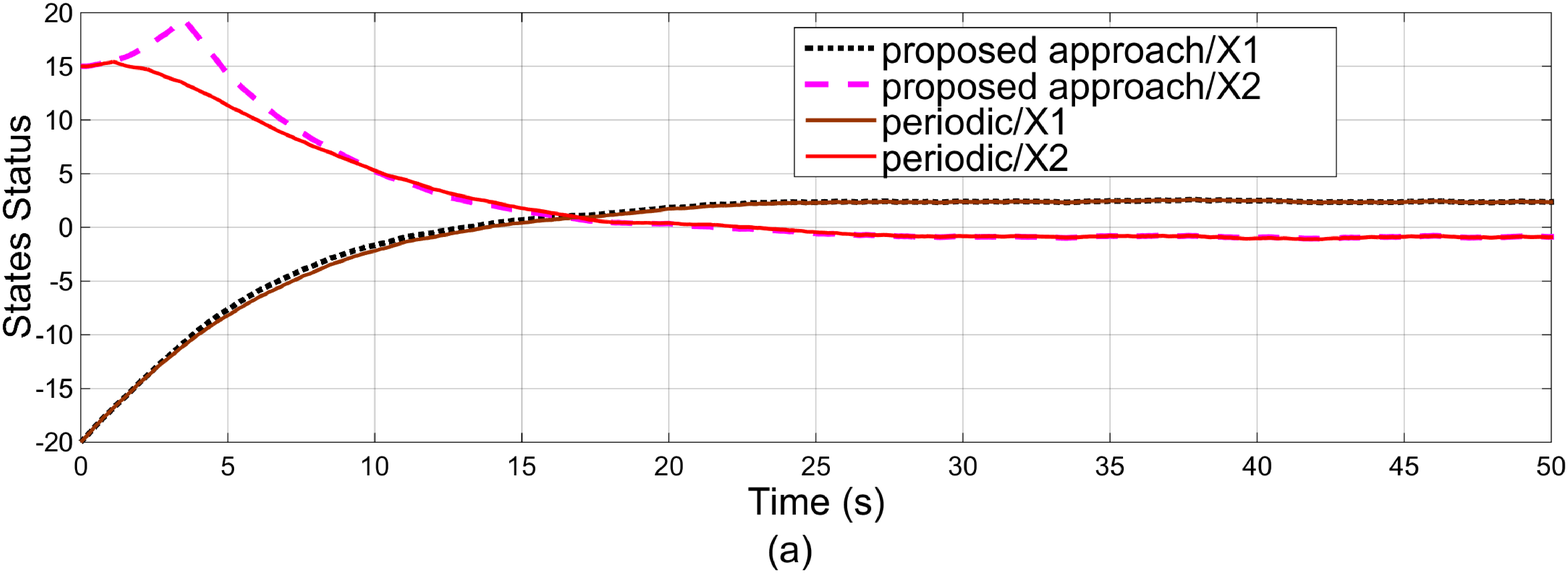}}
 	\centerline{\includegraphics[width=0.5\textwidth]{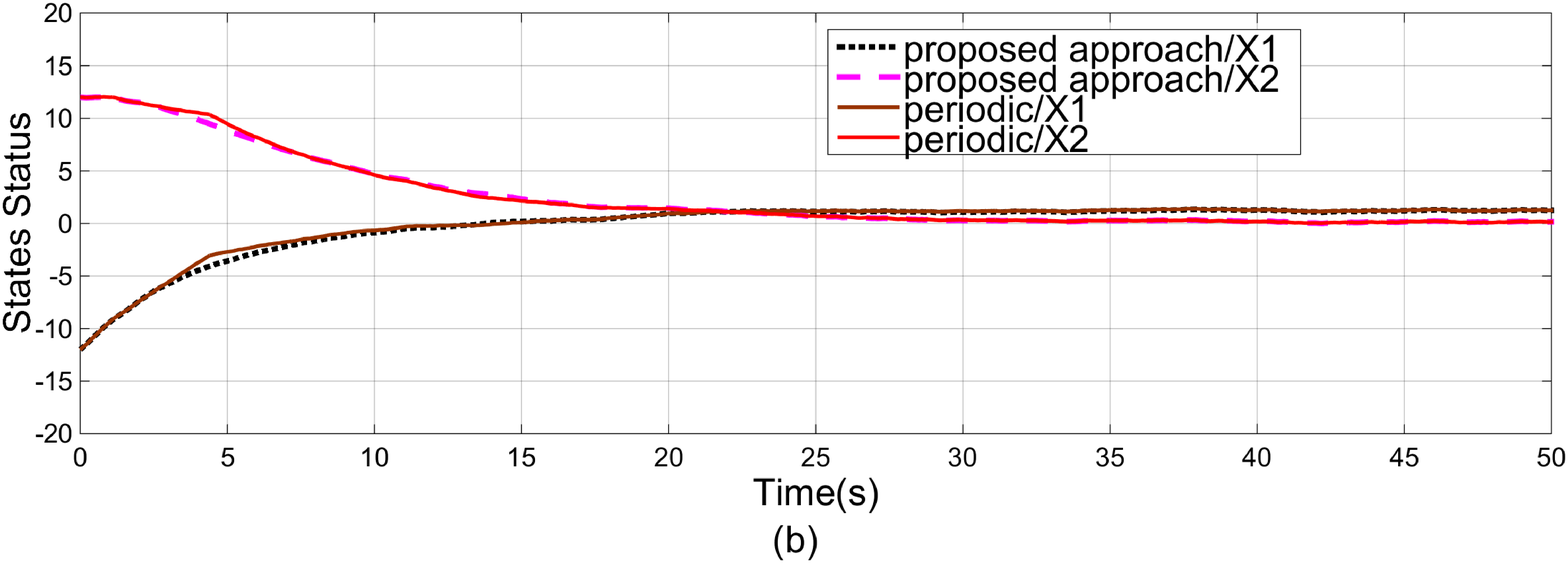}}
 	\centerline{\includegraphics[width=0.5\textwidth]{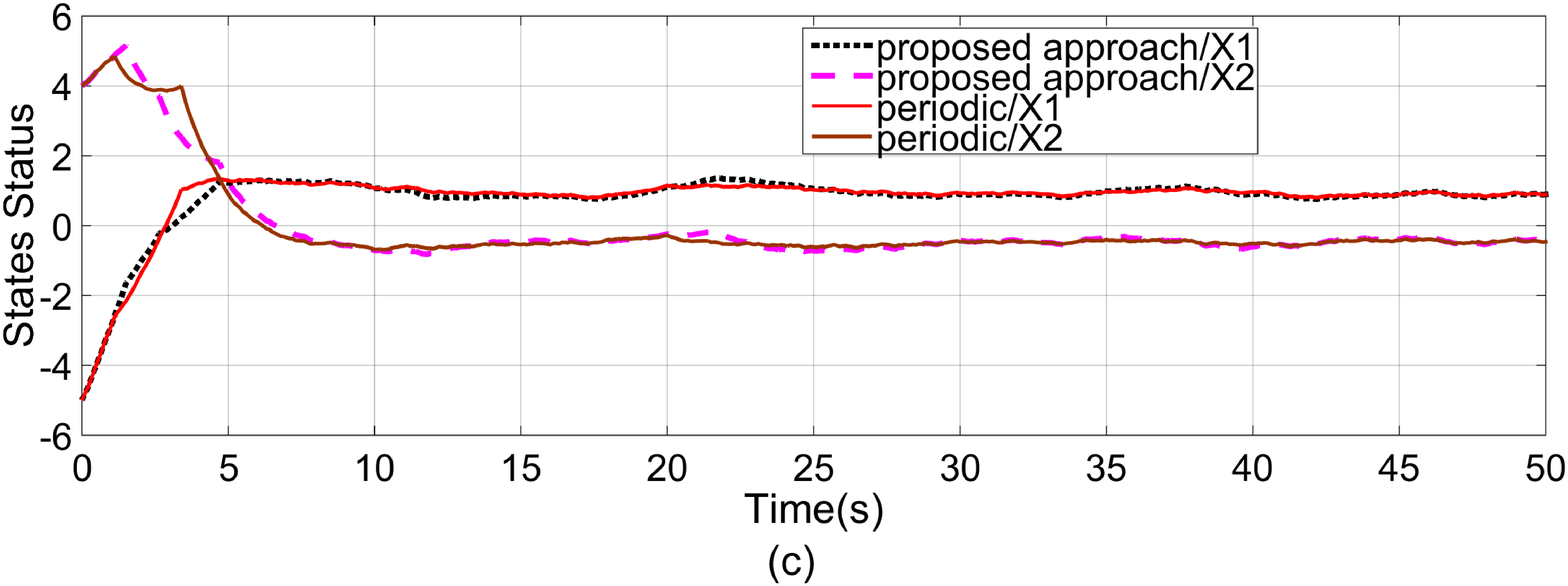}}
 	\label{fig:Plants-Responces}
 	\caption{Analogy between CPS responses of proposed approach and periodic method implementation, (a) First control plant response, (b) Second control plant response, (c) Third control plant response}
 \end{figure}  
 
Figs.4 and 5 illustrate the total power consumption ($P^u_{total}+P^{BS}_{total}$), the power consumption in uplink ($P^u_{total}$) and downlink ($P^u_{total}$) for about 50s simulation time, when periodic method and our proposed approach are applied, respectively. In both Figs.4 and 5, the minimum point of the all power consumption types is taken place at $a=0.5$, when the priority of objective functions ($a$) is equal. Fig.6 clearly illustrates the comparison of the power consumption in our proposed approach with Periodic method Implementation. It is observed that the power consumption in our proposed approach is significantly lower than periodic method implementation.    

\begin{figure}[h!]
	\centerline{\includegraphics[width=0.5\textwidth]{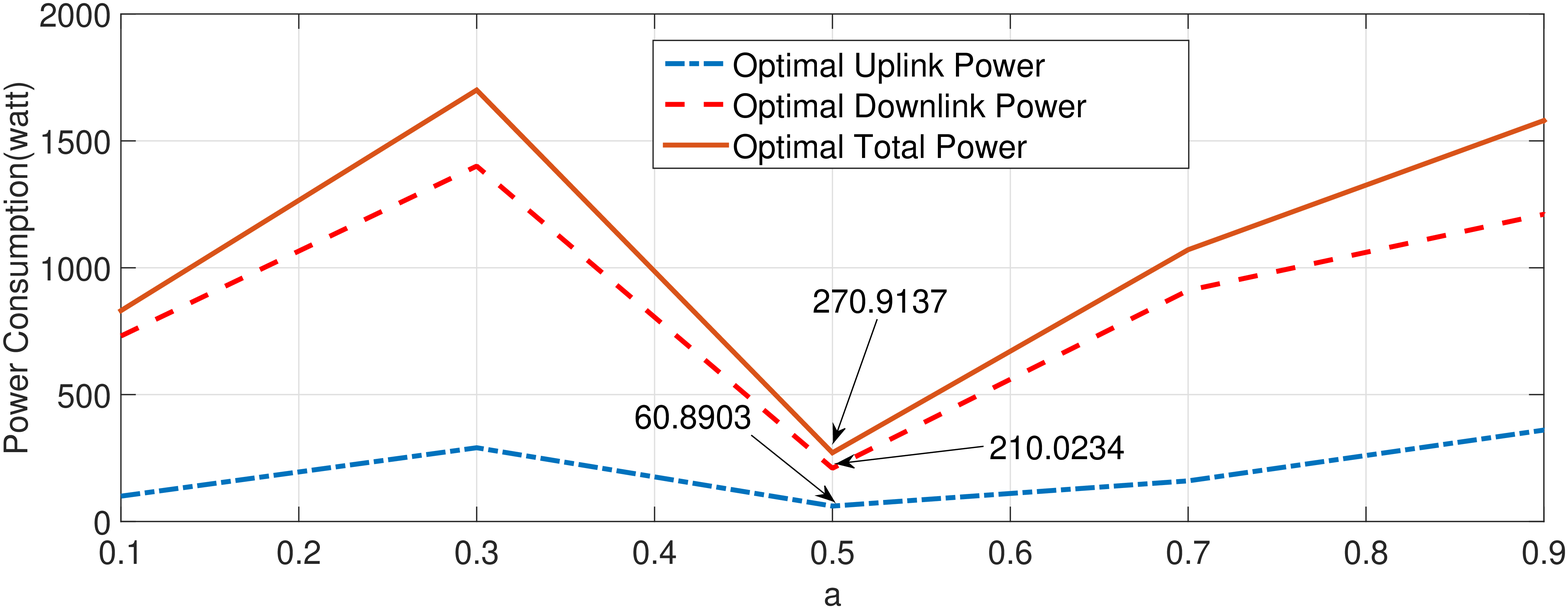}}
	\label{fig:Powerperiodic}
	\caption{Power consumption in periodic method Implementation }
\end{figure}   
 
 \begin{figure}[h!]
 	\centerline{\includegraphics[width=0.5\textwidth]{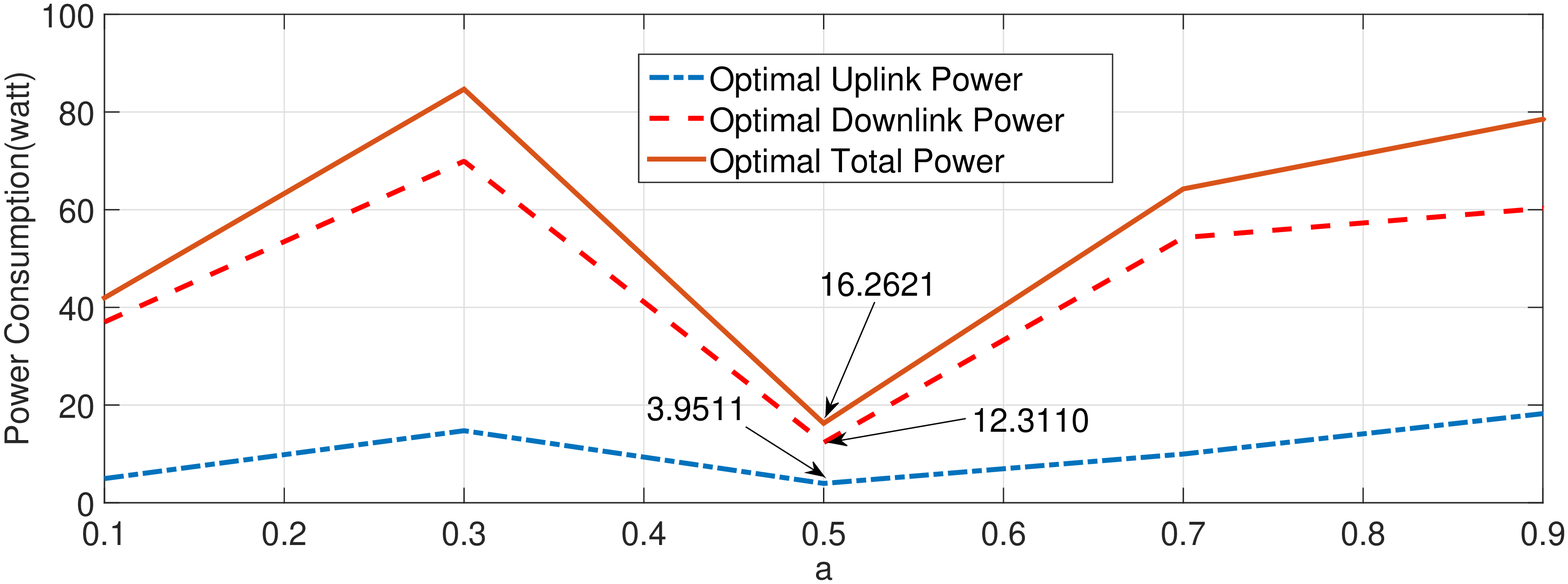}}
 	\label{fig:Powerself}
 	\caption{Power consumption in our proposed approach}
 \end{figure}

 \begin{figure}[h!]
	\centerline{\includegraphics[width=0.5\textwidth]{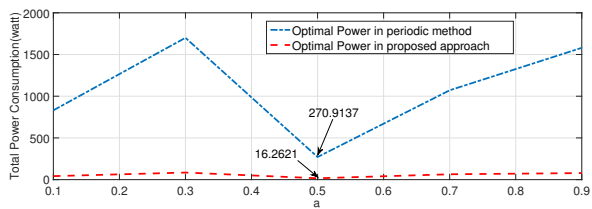}}
	\label{fig:comparepower}
	\caption{Comparison of the total power consumption in our proposed approach with periodic method implementation }
	\end{figure}
 
\section{Conclusion and Future Work}
We have studied the problem of power minimization in the industrial CPS. To realize power minimization in the CPS, the problem of jointly decreasing sampling times in control subsystem and allocating power-efficient radio resources in network subsystem is formulated. The objectives of the problem are determining the next maximum allowable sampling instant of each control plant and power minimization in uplink and downlink, considering the dynamics and desired performance of each control plant, the QoS requirement of RC users, power and sub-carrier constraints. We proposed a novel approach to address this multi objective problem. The proposed approach decomposes the optimization problem into two smaller loosely coupled problems. We showed that our proposed approach considerably decreases power consumption, whereas the stability statuses and control performances of control plants responses are satisfied. In this paper, we conservatively assumed a fixed maximum value for computation delay variable. Considering an actual variable for computation delay is a complicated scenario which will be studied in our future works.   

\IEEEtriggeratref{20}
\bibliographystyle{IEEEtran}
\bibliography{IEEEabrv,Power_efficeint_Sampling_Time_and_Resource_allocation_in_CPS_with_Industrial_Application}

\end{document}